# White Light Generation and Few Cycle Pulse Compression in Cascaded Multipass Cells.

SEMYON GONCHAROV,*[1] KILIAN FRITSCH[1,2], OLEG PRONIN[1]

[1]Helmut-Schmidt-Universität / Universität der Bundeswehr Hamburg, Holstenhofweg 85, D-22043 Hamburg, Germany
[2]n2-Photonics, Hans-Henny-Jahnn-Weg 53, 22085 Hamburg

*Corresponding author: semyon.goncharov@hsu-hh.de

We report supercontinuum generation and pulse compression in two stacked multipass cells based on dielectric mirrors. The 230 fs pulses at 1 MHz containing 12 µJ were compressed by factor 33 down to 7 fs, corresponding to 1.0 GW peak power and overall transmission of 84 %. The source is particularly interesting for such applications as time-resolved ARPES, photoemission electron microscopy, and nonlinear spectroscopy.

Spectral broadening and compression of ultrashort pulses is a widely used technique to achieve high pulse broadening and compression factors and to generate few-cycle electric fields, which are beneficial for a wide range of applications like ultrafast pump-probe spectroscopy [1] and nanoscopy [2], attosecond science [3] and X-ray or XUV sources [4,5]. Lately, a fundamentally new approach for spectral broadening and pulse compression in a quasi-waveguide system, Herriott-type multi-pass cell (MPC), was introduced [6]. Once properly coupled into a cell, a laser beam can propagate distances much longer than the cell's geometrical size. This concept can be applied for spectral broadening and pulse compression when pulses propagate through a nonlinear medium inside the cell multiple times, thus, accumulating sufficient nonlinear phase shift. Herriott cells are less prone to misalignment than conventional fiber-based systems and less susceptible to damage at high peak and average powers. The technique was applied to bulk- and gaseous multipass cells covering an impressive input peak power range from 9 MW [7] to 81 GW [8], respectively. However, only a few demonstrations approached the sub 15 fs pulse duration as it gets increasingly difficult to support the necessary bandwidth and manage the corresponding mirror dispersion. In contrast to fiber waveguides exploiting the total internal reflection, and, thus, in principle, not limited in the spectral bandwidth, the multi-pass cells require mirrors that should have metal or dielectric coatings. Metal coatings have big advantage of being extremely broadband and dispersion-free, however, showing significant absorption losses on the order of a few percent per reflection. Dielectric mirrors can approach one octave [9,10] bandwidth, have losses on the order of 0.1%, and additionally show the oscillations of group delay dispersion (GDD). Importantly, even though fibers can support extreme bandwidths, efficient compression down to a few cycles and below requires excellent dispersion control. This, in turn, requires dispersive mirrors, which can compensate for higher-order dispersion terms. This was, for example, implemented in the light-field synthesizers [11]. In principle, similar concepts having different spectral channels broadened in the multi-pass cell and consequently compressed are possible but complex and cumbersome in implementation. In work [12], the authors recently utilized a gas-filled two-stages multipass system. The first stage implemented dielectrically coated mirrors, resulting in the output pulses of 31 fs. The second stage was based on silver-coated silicon-substrate mirrors and showed a transmission of 82 %. An overall transmission of 78 % and 61 % of energy in the main peak of a 6.9 fs long pulse was demonstrated. Moreover, a 90-fold compression factor of 112 µJ with a pulse duration down to 11 fs was recently achieved in a double-staged, purely bulk-based MPC [13]. In [14], Fritsch et al. showed that the compression of 220 fs pulses down to 16 fs is possible with an overall transmission of 60 % and relatively low input peak power and high repetition rate. These results are summarized in table 1. In this work, we rely on dispersive mirrors covering the bandwidth from 850 to 1300 nm and Argon gas as a nonlinear medium. Thanks to the dielectric mirrors, these kinds of cells can be average power scalable without excessive engineering efforts and, importantly, show excellent pulse quality with 80 % energy preserved in the main peak and high overall transmission of 84 %. Moreover, we experimentally show that with proper intra-cell dispersion management, an MPC can operate in normal, zero- or anomalous dispersion regimes. The latter is particularly interesting due to different highly pronounced nonlinear phenomena like self-compression and white-light generation easily observed in the fibers [15]. MPCs are very robust with regard to misalignment-triggered damage and are suitable for high power operation, maintaining average power scalability [8,12,16] and compact portable form-factor. The abovementioned merits drastically mitigate requirements for a wide range of applications, such as high-repetition rate extreme ultraviolet sources via high harmonic generation (HHG) in noble gases [4] for ARPES [17] or XUV-frequency comb spectroscopy. Also, it provides new possibilities for multiphoton microscopy in the near-infrared region [18], where peak powers are rather low, and the optimal repetition rates are in the range of a few MHz. One such application could be the hyperspectral multiphoton microscopy [19], which allows for the simultaneous excitation of different chromophores with a single ultra-broadband laser source.

**Table 1.** A summary of multipass systems with sub-20 fs output pulses; input and output peak powers $P_{pk}$, pulse durations $\tau$, average powers $P_{avg}$, overall transmission calculated as the multiplication of individual stages transmission, repetition rate $f_{rep}$.

|  | Units | This work | [14] | [12] | [13] | [20] | [21] |
|---|---|---|---|---|---|---|---|
| $P_{pk\ input}$ | MW | 48 | 14 | 4700 | 105 | 21 | 1567 |
| $P_{pk\ output}$ | MW | 1025 | 60 | 60000† | - | - | - |
| $\tau_{input}$ | fs | 230 | 220 | 200 | 1000 | 265 | 1200 |
| $\tau_{output}$ | fs | 7 | 16 | 7 | 11 | 18 | 13 |
| $P_{avg,input}$ | W | 12 | 100 | 500 | 112 | 100 | 200* |
| $P_{avg,output}$ | W | 10 | 60 | 388 | 15 | 61 | 37* |
| Transm. | % | 84 | 60 | 78 | 47 | 61 | 37 |
| $f_{rep}$ | MHz | 1 | 28 | 0.5 | 1 | 16 | 0.1 |

\* in-burst average power
† power-attenuated beam compressed to 2.3 GW

In contrast to the experiments summarized in Table 1, we relied on the combination of the gas medium and dispersive dielectric coatings in order to fine-tune the cell's dispersion precisely and, this way, reach optimal broadening and white-light generation.

Moreover, it was possible to controllably go through positive, zero, and overall negative dispersion. Thus, we operated the system both in positive and negative dispersion regimes and showed that reaching a few-cycle pulse durations accompanying a high temporal compression factor of 33 with fs-long seed pulses is feasible in simple and robust geometry.

The experimental setup is shown in Fig. 1. The driving laser represented a commercially available Pharos laser delivering 230 fs long pulses at 1 MHz containing 12 µJ energy resulting in 12 W average power. This laser output was chosen to mimic the output peak power of our thin-disk oscillator [22] and be able to use a similar multipass concept for this oscillator. The laser output was mode-matched to the first multi-pass cell eigenmode. The first Herriott cell consisted of two highly reflective mirrors (150 mm, and 100 mm radius of curvature, respectively) separated by 190 mm and a 3 mm-thick AR-coated fused silica placed approximately 10 cm away from the input mirror. The 30 passes through the fused silica plate yielded a total B-integral of 12.9 rad, corresponding to 0.43 rad per pass. This multipass cell operated in an ambient air environment. The output pulses were compressed to 40 fs (Fig. 2) with 9 bounces off dispersive mirrors, providing each -400 fs$^2$ of GDD per bounce. The compression was verified by the commercial SPIDER device (APE GmbH). The transmission of the first stage remained at 94 % and is mainly attributed to losses in the AR-coated fused silica.

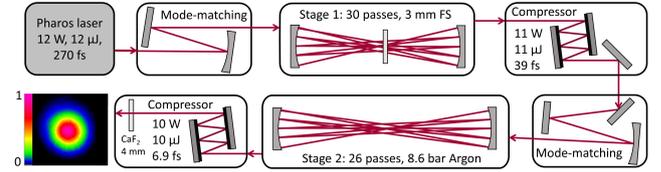

**Fig. 1.** Schematic setup of the nonlinear broadening and pulse compression.

Additionally, we characterized spatio-spectral homogeneity by measuring spectra in different positions of the compressed beam. The beam was coupled into a spectrometer via a 200 um-core-size multimode fiber. The overlap parameter [7] was used to quantify the measurement (Fig. 3a). The result showed a good overlap of > 99 % at the center of the beam and going down to 90 % at the edges (defined as $1/e^2$) while the weighted mean values lie around $V_x = 99$ % and $V_y = 99$ %, respectively.

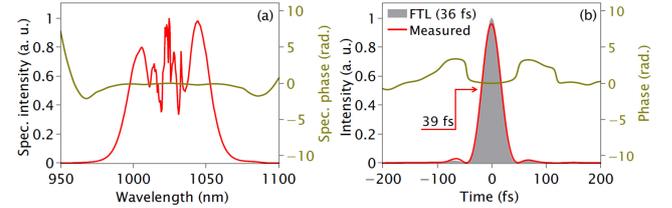

**Fig. 2.** Temporal characterization of the stage 1 output. (a) Spectral phase and intensity retrieved with a commercial SPIDER device. (b) Temporal phase and intensity compared to Fourier transform limit of the spectrum. The main peak included 96 % of the pulse energy.

The output beam had an excellent quality and was measured in accordance with ISO 11146 to be $M^2 = 1.2 \times 1.2$ (Fig. 3b). Both values are comparable to the driving laser beam quality $M^2 \leq 1.2$.

In the following step, the compressed pulses were coupled via a mode-matching mirror to the second stage. The second Herriott cell included a complementary pair of dispersive mirrors with a 200 mm radius of curvature separated by 380 mm. The assembly was placed in a monolithic aluminum housing filled with Argon. The housing had an AR-coated fused silica input window. The configuration of the cell included 26 passes through the gas volume. The group velocity dispersion of Argon was considered 0.015 fs$^2$/mm [23]. The cell mirrors were designed to provide approximately -100 fs$^2$ after two bounces around 1030 nm. Thus, the cell can operate in normal, zero-, or anomalous dispersive regimes depending on gas pressure. Based on the numerical estimations, a pressure of ~8.5 bar was considered a transitional value between the regimes. Firstly, the second stage was tuned in a slightly positive dispersive regime corresponding to 8.6 bar of Argon pressure. That allowed us to fully utilize the entire bandwidth of the broadband mirrors (850 – 1300 nm), barely sacrificing the overall throughput. The second stage output spectrum is shown in Fig. 4.

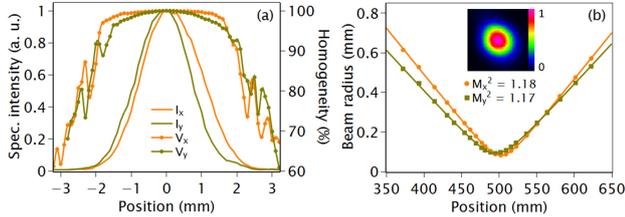

**Fig. 3.** Characterization of the stage 1 output. (a) Spatio-spectral homogeneity V and relative intensity I for tangential and sagittal planes. (b) Beam quality measurement $M^2$ according to ISO 11146. The dots and the lines represent experimental data and fit, respectively. The beam in the focal plane is depicted in the inset.

The pulse duration and the phase information were retrieved with the commercial SPIDER device (see Fig. 4). Based on the retrieval, the main peak contained 80 % of energy or ~1.0 GW peak power. This corresponds to factor 33 of temporal compression and factor 21 of peak power increase. The second stage throughput in the positive dispersion regime was 89 % and was essentially defined by the cell mirror coatings, while the overall transmission of both stages remained 84 %. The output spectrum was measured by an optical spectrum analyzer from Ando AQ6317B (Fig. 4a). The Fourier transform limit of 6.5 fs was calculated from the measured spectrum. The output pulses were compressed with 4 bounces off dispersive mirrors (4 x -45 fs$^2$) and a 4 mm thick CaF$_2$ window down to 7 fs.

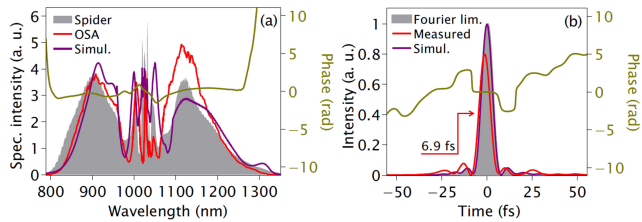

**Fig. 4.** Temporal characterization of the stage 2 output. (a) Spectral phase (olive) and intensity (grey) were retrieved with a commercial SPIDER device. The spectrum was additionally measured with an optical spectrum analyzer (OSA). The purple curve represents the simulated output spectrum. (b) Temporal phase and intensity are compared to the spectrum's Fourier transform limit (6.5 fs). The main peak included 80 % of the pulse energy.

The nonlinear phase shift/B-integral in the second stage was estimated as 0.74 rad per pass. No significant deterioration of the output beam was observed (see Fig. 1). However, the sensitivity of our CCD sensor spans the range of 320 – 1100 nm, which allowed us to verify the beam quality after the second stage only within this spectral range (See Fig. 5c). The beam quality parameter was measured to be $M^2$ = 1.2 x 1.2, thus showing the absence of beam degradation when propagating through cascaded multipass stages.

The spatio-spectral homogeneity of the output beam was characterized following the procedure of the measurement of the first stage output (Fig. 5a). A perfect spectral overlap of > 99 % was measured in the central part (defined as $1/e^2$) of the beam while going down to 90 % at the edges. The weighted average values of overlap factors were $V_x$ = 99 % and $V_y$ = 99 %, respectively. The results indicated an excellent spectral content enclosure over the beam area.

The long-term power stability measurement (Fig. 5b) was performed over 13 hours of continuous operation. The system ran stably within the measurement period without a significant drop in output power.

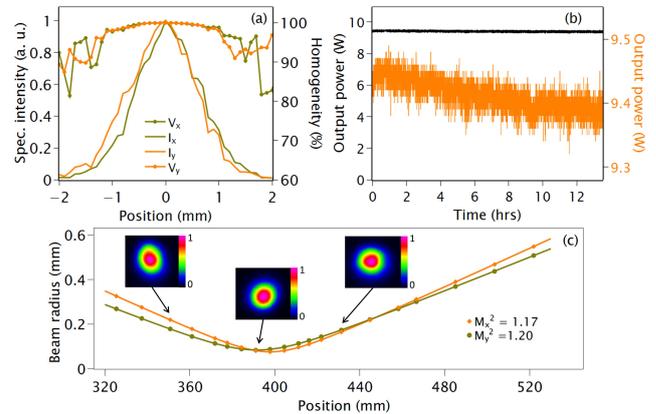

**Fig. 5.** Characterization of the stage 2 output. (a) Spatio-spectral homogeneity V and relative intensity I for tangential and sagittal planes. (b) Long-term measurement of the output power. (c) Measurement of the beam quality $M^2$. The central wavelength of 878 nm is an average weighted according to the sensitivity of the CCD sensor. The insets represent the beam profile at different positions within the caustics.

As a next step, we fine-tuned the gas pressure down to 8 bar to reach a net anomalous dispersive regime when the positive dispersion of Argon was slightly overcompensated by the negative dispersion of the cell mirrors. This led to a partial self-compression of the input pulses inside the cell and allowed us to maximize the broadening in the current cell configuration. Mainly, the output spectrum covered the full octave in the 700 – 1450 nm range. However, this compromised the overall throughput to < 80 % due to the high linear losses of the broadband mirrors beyond their working region. This experiment shows that the octave-spanning bandwidth and, ultimately, white light generation is feasible with the multipass cells.

We carried out simple 1D simulations of pulse propagation through a quasi-waveguide filled with a nonlinear material to support the experimental results (see Fig. 4). Our simulations comprised numerical solving of the nonlinear Schrödinger equation using an open-source software PyNLO [24]. The grid consisted of $2^{14}$ points spanning over a 20 ps time window. Each iteration of the algorithm included 50 steps of propagation through the cell in one direction, considering the positive dispersion of Argon and the form of the caustics. Afterward, a negative dispersion of the cell mirror was applied to the pulse. Additionally, the transmission curves of the cell mirrors were considered. This procedure allowed us to match the conditions of the experiment closely. The experimental results are in good agreement with the numerical simulations. Considering theoretically flat dispersive (with GDD of -50 fs$^2$) and transmission curves (~0.1 %) of the broadband mirrors spanning over one octave [25], we extended the simulations beyond experimental conditions to show the perspective of the spectral broadening and pulse compression in multipass cell geometry. As shown in Fig. 6, an efficient (> 90 % throughput) entire octave span is achievable, paving the path to a compressible sub-2 optical cycle regime with an FTL < 5 fs.

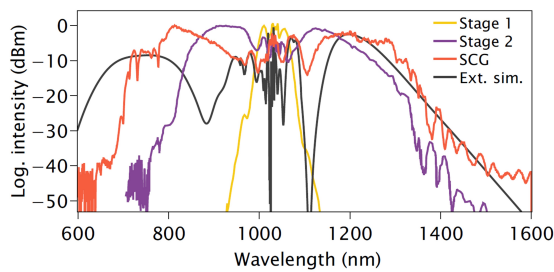

**Fig. 6.** Output spectra in the setup. The curves are measured with an optical spectrum analyzer. Yellow and purple curves correspond to the output of stage 1 and stage 2, respectively. The orange curve is attributed to supercontinuum generation (SCG) when stage 2 operated in an anomalous dispersion regime. The grey curve is a simulated SCG assuming spectrally extended properties of cell mirrors.

In conclusion, we demonstrated a simple and compact dual-stage Herriott-type multipass system, which compressed 230 fs input pulses to 7 fs with an overall transmission of 84 % and maintained 80 % energy in the main peak. Additionally, supercontinuum generation was demonstrated while operating the second stage in a net anomalous dispersion regime. The results clearly show the possibility of achieving an efficient sub-2 optical cycle regime with FTL < 5 fs. The setup relying on the all-dielectrically coated mirrors and gas as a nonlinear medium ideally suits the demand of the spectral broadening and compression of high average and peak power Yb-based lasers. The laser system is an ideal high repetition rate driver for HHG-based sources and ultrafast pump-probe experiments.

**Acknowledgments.** Considering the difficulty in setting up the new professorship in combination with typical university bureaucratic procedures, we would like to sincerely acknowledge a few facilitators: A. Borchers, D. Kiesewetter, and A. Puckhaber. Additionally, we thank Christian Franke for designing the optomechanical components and cell housings.

**Disclosures.** OP and KF declare a conflict of interest as co-founders of n2-Photonics.

**Data availability.** Data underlying the results presented in this paper are available on request.